# Quantifying the Impact of Epigallocatechin Gallate and Piperine on D. tigrina Regeneration

1st Vinil Raj Polepalli
*High Technology High School*
Manalapan, US
vinil.polepalli@gmail.com

2st Pranav Samineni
*Harvard University*
Boston, MA
psamineni@hms.harvard.edu

*Abstract*—The increasing global cancer burden necessitates exploration of effective and affordable treatments. Epigallocatechin gallate (EGCG), a green tea catechin, and piperine, a black pepper alkaloid, have demonstrated promising anti-cancer properties. Leveraging the regenerative capacity of Dugesia tigrina (planaria) and their neoblast stem cells as a model for cancer growth, this study investigated the combined effects of EGCG and piperine on cell proliferation. Planaria were exposed to varying concentrations of EGCG and piperine over seven days, with growth changes recorded and compared to a negative control group. Initial trials identified optimal concentrations for growth inhibition, subsequently validated in a second trial using combined EGCG and piperine treatment. Statistical analysis revealed significant differences in growth across all experimental groups ($p < 0.05$), indicating a synergistic effect of EGCG and piperine in limiting planarian growth. These findings suggest the potential of combined EGCG and piperine therapy for cancer and other proliferative diseases like keloids and psoriatic arthritis, warranting further investigation into clinical applications.

I. INTRODUCTION

Cancer treatments have been a major focus of the medical field for decades. Some common methods that combat cancer include radiation, chemotherapy, and hormonal therapy [10]. However, despite undergoing rigorous clinical trials, many of these treatments still have flaws. In 2019, the survival rate for individuals diagnosed with various cancers between 0 and <10 years after diagnosis was approximately 55% [10]. In addition, 49% of cancer patients experienced financial hardship [8]. Today, there is a need for a highly effective and affordable treatment. Compounds such as epigallocatechin gallate (EGCG) and piperine have shown promise as potential cancer treatments that fit these constraints [8].

Epigallocatechin gallate (EGCG), a green tea polyphenol, has been shown in vitro to inhibit the exponential growth of human cancer cells and activate the ERK pathway—an effect that can both promote oncogenesis and induce apoptosis—though it has yet to undergo large-scale clinical trials to confirm broader efficacy [1][5]. Notably, EGCG's anticancer activity is enhanced when combined with other natural compounds such as resveratrol and curcumin [7], and its low cost makes it an attractive candidate for future cancer therapies [6]. Similarly, piperine—the bioactive alkaloid from black pepper (Piper nigrum)—exhibits anti-inflammatory, antioxidant, and cancer‑preventive properties, with studies indicating it may disrupt the Wnt/β-catenin signaling pathway implicated in cell proliferation and survival [4]. While both EGCG and piperine show promise as complementary or standalone anticancer agents, further research is needed to elucidate their mechanisms, optimize combinations, and validate their therapeutic potential.

Conversely, studies regarding piperine are relatively limited. Choosing a specific outcome, like a statistically higher or lower amount of cell growth, is difficult, as different amounts of EGCG may either promote cell growth or hinder stem cell differentiation. However, when combinations of EGCG and piperine are administered, significant differences in cell growth are expected due to the effects on the ERK pathway, inhibition of cancer growth, and cell proliferation properties [5].

Planaria have been selected as the model organisms for this study due to their genetic pathways and their characteristics similar to those of humans. The similarities include comparable patterns of cell growth, differentiation by stem cells, and programmed apoptosis. By examining the growth changes in planaria when exposed to EGCG and piperine, this study aimed to show how planaria's growth can change when EGCG and piperine are induced. The several experimental advantages of planaria are their transparent body, small size, and short lifespan, making them a resource-efficient choice for research purposes [8].

With planarian stem cells, there are signaling pathways like Wnt, Hedgehog, and Notch that cause the regeneration to occur (Sánchez Alvarado & Kang, 2005). These pathways are shared between cancer cells and planarian stem cells. The EGCG and Piperine have shown signs of inhibiting these pathways when involved with cancer cells. Planarian stem cells are also a model for cancer cell anti-apoptosis. In planarian regeneration, only damaged cells are regenerated. When exposed to EGCG and piperine, regeneration may be impaired due to apoptosis-inducing (via ERK and/or Wnt/β-catenin pathways) or regeneration-blocking (via Wnt,

Hedgehog, and/or Notch) pathways. Regeneration in this model is analogous to the uncontrolled proliferation of cancer cells, and thus, agents like EGCG or piperine may also inhibit cancer growth.

This study consisted of three distinct investigations. The first study aimed to examine the impact of increasing concentrations of EGCG on the regeneration rate of planaria. It was hypothesized with the knowledge of EGCG's cancer-preventive properties, varying concentrations of EGCG could yield a statistically significant effect on planaria's regeneration rate [8]. The second study was similar to the first, but with piperine as the independent variable. It was hypothesized that, knowing piperine's cancer-preventive properties, varying treatment could yield a statistically significant effect on planaria's regeneration rate. The third investigation explores the combined effects of EGCG and piperine on planaria by simultaneously increasing the concentrations of the two compounds. The independent variables in the study were EGCG and piperine, while the dependent variable was the growth of planaria. It was hypothesized that, knowing EGCG and piperine both have positive cancer-preventive properties, combined treatment could prevent the cell proliferation of the planarian stem cells. Because of these effects, it was predicted that EGCG and piperine would prevent the growth of the planaria.

This study aimed to identify the effects of EGCG and piperine on Dugesia Tigrina stem cells, aiming to improve human cancer treatments, especially in terms of effectiveness, safety, and cost. By exploring the cancer-inhibiting properties of piperine and EGCG in planaria, this study hopes to gain insights that could be translated into human cancer treatment strategies in future studies [8].

## II. MATERIALS AND METHODOLOGY

### A. Materials, Equipment, and Facilities

The study used Dugesia tigrina planaria from High Technology High School, housed in 50 mm Petri dishes and maintained in Poland Spring water at ~20 °C. Treatments were prepared by dissolving EGCG (99.87% purity, 50 mg total) and piperine (98.88% purity, 5 g total) in the same water. Planaria (5 per dish) were bisected with a scalpel, then imaged under a dissecting light microscope (HTHS) and with an iPhone 12; ImageJ on a 2023 MacBook M2 Air quantified regeneration metrics. Animals were fed 1 cm³ beef liver chunks every 7 days (non-regenerating) or 10 days (regenerating), and all dishes were kept on the same lab bench to standardize conditions.

### B. EGCG and Piperine: Experimental Design

In this study, EGCG was utilized as a compound with a solubility of at least 5 mg/mL, indicating its relative solubility in water. The materials and equipment used to make the EGCG solution include a beaker of water, a spoon, a micropipette, EGCG powder, a precision scale, a petri dish, and a measuring tray (see Figure 1). The initial step involved filling the beaker with 100 mL of water. Subsequently, 10 mg of EGCG was added to the solution. The study used serial dilution as per Figure 1 to achieve the desired concentration from the initial concentration of 218.164 μmol. When the desired concentration was achieved, the solution was then added to a petri dish to achieve the desired volume of the petri dish, 15 mL. This petri dish was appropriately labeled to indicate the presence of EGCG in the planaria environment. The regenerating planarians were then added to this Petri dish.

For piperine, a similar procedure was used. However, due to piperine's low solubility in water, 40mg/L, changes had to be made. The stock solution was 100 mL of water mixed with 1 gram of piperine. After that, the solution was filtered to remove any excess solid piperine. The concentration of this solution was 140.184 μmol. Besides that, the process for preparing the piperine solution followed the same steps as described earlier. After making the solution, it was added to a labeled petri dish, indicating the presence of EGCG as well as the concentration.

When preparing the EGCG and piperine solution, both EGCG and piperine acted as solutes in the same solution. The required amount to achieve the desired concentration of EGCG and Piperine from both stock solutions was added to the petri dish. It is important to note that the petri dish should remain empty at this stage, as the planaria was introduced subsequently during the experimental setup.

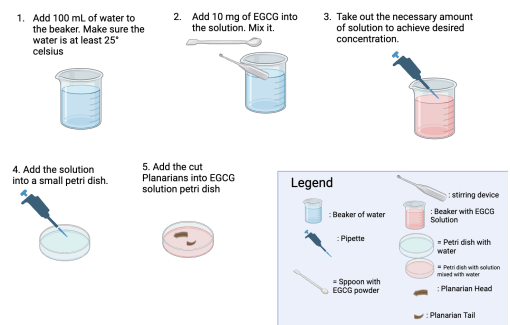

**Figure 1:** Serial Dilution and Exposure of Planarians to EGCG Solution

### C. Planaria Conditioning and Experimental Readiness

The overall experimental procedure consisted of two key stages: a preliminary growing period and the subsequent experimental phase involving all the designated groups. The primary objective of the initial growing stage was to ensure the optimal health and condition of the planaria, thus minimizing any potential mortality resulting from pre-existing health issues, as well as increasing the number of planaria that could be tested. This phase was independent of the experimental groups and aimed to establish a baseline health status for the planaria population. It is noteworthy that the planaria used during the growing period were subsequently employed in the experimental groups, ensuring consistency and comparability across the study. Planaria were cut and subsequently regenerated to increase the population size.

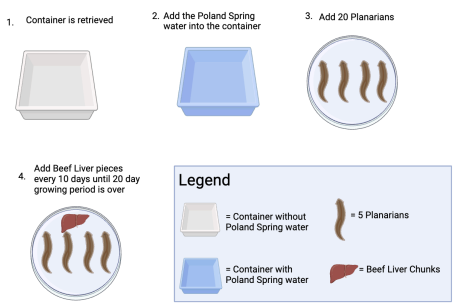

**Figure 2:** Setup for Planarian Growth and Nutritional Conditioning

The growing period was conducted using a large plastic container acquired from HTHS, which was first filled with Poland Spring water before introducing the planaria. Specifically, all the planarians used in this experiment were placed in the petri dish. As illustrated in Figure 2, beef liver chunks were added to the container every ten days, serving as their food source during this period. The combined feeding and growing phase spanned 30 days. To ensure the accuracy of subsequent experimental measurements, any deceased planaria were promptly removed from the container. In between the 10-day feeding periods, the planaria were cut, as seen in Figure 3, and regenerated.

For the organization of planaria into their respective experimental groups, the planaria were cut into two pieces, namely a tailpiece and a headpiece. The cutting process involved a 200 mm petri dish filled with ice, a scalpel, and the experimental petri dishes. As seen in Figure 3, the planaria were placed on the ice-filled petri dish and were subsequently cut. Following the cutting procedure, the planaria fragments were transferred to a separate Petri dish containing the Poland Spring water solution. A transfer pipette was used to move the planaria pieces carefully to their assigned experimental group. The four designated experimental groups include an EGCG solution, a piperine solution, a control group with only Poland Spring water, and a combined EGCG and piperine solution. The combined EGCG and piperine groups were tested after all the trials were complete for the control, EGCG-only, and piperine-only groups. The molarities of the solutions were based on the molarities of the EGCG and Piperine groups that performed the best. Subsequently, the planaria pieces were cultivated for two weeks and fed with beef liver chunks. Photographs were taken every two days for seven days to document the experiment's progression.

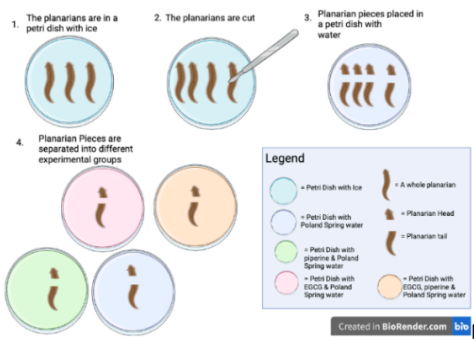

**Figure 3:** Preparation of Planaria for Experimental Group Assignment

### D. Recording Observations and Analyzing Data

Images were captured daily for four weeks by photographing planaria under a dissecting microscope (HTHS) with an iPhone 12, then preprocessed to enhance contrast and remove background noise. Segmentation isolated each worm, and feature extraction quantified length, width, area, and perimeter, while tracking algorithms followed individual specimens over time. Measurements were converted to percent growth, from which growth rates and regenerated tissue percentages were calculated. Descriptive statistics summarized these metrics, and group differences were assessed by ANOVA, followed by pairwise t-tests comparing each treatment to the control.

### III. RESULTS

This study examined the effects of EGCG and Piperine on the regeneration rate of Dugesia tigrina for seven days; data were recorded every two days after the initiation of the experiment. The planaria were measured, and the data were stored in the raw data table. There was a total of 36 planaria for each experimental group. The data were then converted to percent change relative to the day 1 data. This data were stored in calculated data. The mean, standard deviation, variance, and number of trials for all experimental groups were stored in the summative data table. All experimental groups were compared with an initial ANOVA test and a T-test. The data found from these statistical tests were stored in their respective data tables.

TABLE I. DAY 7 REGENERATION (% OF DAY 1) IN D. TIGRINA WITH EGCG/PIPERINE

| The Summative Data Table for Regenerating Dugesia Tigrina by Varying Concentrations of EGCG and/or Piperine Growth for Day 7 (% of Day 1) | | | |
|---|---|---|---|
| *Experimental Group* | *Mean* | *Standard Deviation* | *Variance* |
| Control | 35.719 | 4.258 | 72.519 |
| 25 µmol EGCG | 3.227 | 1.952 | 30.481 |
| 50 µmol EGCG | -15.367 | 7.310 | 213.735 |
| 100 µmol EGCG | -12.916 | 5.309 | 112.754 |
| 2.5 µmol Piperine | 1.138 | 2.223 | 39.540 |
| 5 µmol Piperine | 4.267 | 2.708 | 29.320 |
| 10 µmol piperine | -14.444 | 3.338 | 44.573 |
| 12.5 µmol EGCG and 2.5 µmol Piperine | -1.373 | 3.181 | 40.470 |
| 6.25 µmol EGCG and 3.75 µmol Piperine | 25.508 | 3.159 | 39.920 |
| 18.75 µmol EGCG and 1.25 µmol Piperine | -11.484 | 6.503 | 169.143 |

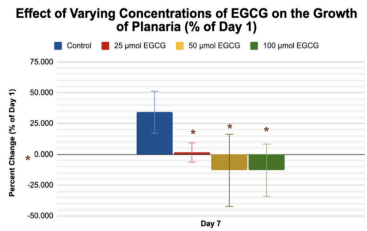

**Figure 4:** Effect of Varying Concentrations of EGCG on the Growth of Planaria (% of Day 1)

This chart shows the effect of varying concentrations of EGCG (25 μmol, 50 μmol, 100 μmol) on the Growth of Planaria (% of Day 1). The error bars represent their experimental groups' mean growth (+- the standard deviation). All of the experimental groups in this graph were statistically significant compared to the control using an ANOVA and a T-test. The p-value of the entire group via the ANOVA test was 0.00000415040. The p-value is less than the alpha value, 0.05, indicating statistical significance. The p-value of the 25 μmol EGCG group, via a T-Test comparing this group and the control, was 0.00005910. The p-value is less than the alpha value, 0.05, indicating statistical significance. The p-value of the 50 μmol EGCG group, via a T-Test comparing this group and the control, was 0.00013542. The p-value is less than the alpha value, 0.05, indicating statistical significance. The p-value of the 100 μmol EGCG group, via a T-Test comparing this group and the control, was 0.00013542. The p-value is less than the alpha value, 0.05, indicating statistical significance. *: Indicates statistical significance.

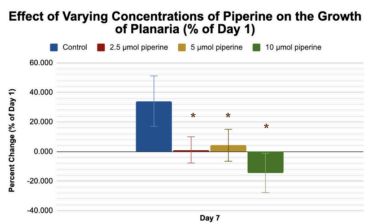

**Figure 5:** Effect of Varying Concentrations of EGCG on the Growth of Planaria (% of Day 1)

This chart shows the effect of varying concentrations of piperine (2.5 μmol, 5 μmol, 10 μmol) on the Growth of Planaria (% of Day 1). The bars represent their experimental groups' mean growth (+- the standard deviation). All of the experimental groups in this graph were statistically significant compared to the control using an ANOVA and a T-test. The p-value of the entire group via the ANOVA test was 0.00000004429. The p-value is less than the alpha value, 0.05, indicating statistical significance. The p-value of the 2.5 μmol piperine group, via a T-Test comparing this group and the control, was 0.00004538. The p-value is less than the alpha value, 0.05, indicating statistical significance. The p-value of the 5 μmol piperine group, via a T-Test comparing this group and the control, was 0.00008678. The p-value is less than the alpha value, 0.05, indicating statistical significance. The p-value of the 100 μmol EGCG group, via a T-Test comparing this group and the control, was 0.00000937. The p-value is less than the alpha value, 0.05, indicating statistical significance. *: indicates statistical significance.

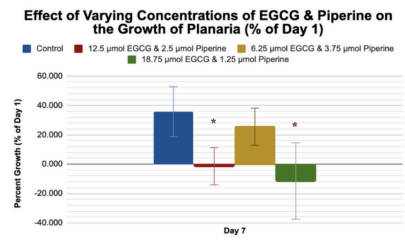

**Figure 6:** Effect of Varying Concentrations of EGCG and Piperine on the Growth of Planaria (% of Day 1)

This chart shows the effect of varying concentrations of EGCG and Piperine (12.5 μmol for EGCG and 2.5 μmol for piperine, 6.25 μmol EGCG and 3.75 μmol piperine, 10 μmol) on the Growth of Planaria (% of Day 1). The bars represent their respective experimental groups' mean growth (+- the standard deviation). All of the experimental groups in this graph were statistically significant compared to the control using an ANOVA and a T-test. The p-value of the entire group via the ANOVA test was 0.00000000642. The p-value is less than the alpha value, 0.05, indicating statistical significance. The p-value of the 12.5 μmol EGCG and 2.5 μmol Piperine group, via a T-Test comparing this group and the control, was 0.00003055. The p-value is less than the alpha value, 0.05, indicating statistical significance. The p-value of the 6.25 μmol EGCG and 3.75 μmol Piperine, via a T-Test comparing this group and the control, was 0.03952665. The p-value is less than the alpha value, 0.05, indicating statistical significance. The p-value of the 18.75 μmol EGCG and 1.25 μmol Piperine group, via a T-Test comparing this group and the control, was 0.00000109. The p-value is less than the alpha value, 0.05, indicating statistical significance. *: Indicates statistical significance.

### IV. DISCUSSION AND CONCLUSION

Combined EGCG and piperine treatments significantly inhibited planarian regeneration at all tested concentrations (Figures 4–6), as confirmed by ANOVA and subsequent one-tailed, unequal-variance t-tests versus control.

Figure 4 shows that the 25 μmol EGCG had the lowest percent change in the growth of planaria compared to the other groups, with an average percent change of 3.227%. It was also observed that this group's standard deviation and variance were lower than the other experimental groups seen in this figure. Based on the data, it was found that 25 μmol EGCG was the ideal concentration to limit any growth of the planaria. Additionally, with the 50 μmol EGCG and 100 μmol EGCG groups, the planaria experienced an average decrease in growth of greater than 10 percent. This may be attributed to extreme environmental stress due to a high concentration of positive apoptosis properties. An important note for the shrinkage is that no planaria died; it gradually reduced in size. This pattern was seen across all of the trials containing EGCG. The error bars in this figure represent one standard deviation.

Figure 5 shows that the 2.5 μmol piperine had the lowest percent change in the growth of planaria compared to the other experimental groups. The 2.5 μmol piperine exhibited an average percent change of 1.138 %. As for the rest of the experimental groups, the 10 μmol piperine group exhibited an average decrease in length of more than 10%. This can be attributed to either extreme environmental stress or positive apoptotic properties. The 5 µmol piperine did exhibit an average percent change of 4.267%. However, the standard deviation of this group was greater than the 2.5 µmol group. Based on the data, a concentration between 2.5 µmol and 5 µmol would be ideal to limit planaria growth. The error bars in this figure represent one standard deviation.

Based on the data from the piperine and EGCG trials, a new set of trials was conducted that had varying concentrations of both EGCG and piperine in the springwater environment of the planaria. In Figure 6, the group with the least percent change was the 12.5 µmol EGCG and 2.5 µmol piperine. This group had an average percent change of -1.373. Overall, a pattern was found in all the groups. When the EGCG ratio to piperine was low, this led to positive growth of the planaria, but it was still statistically significant when compared. However, when the EGCG ratio to piperine was higher, this led to a shrinkage in the planaria. In a previous study, piperine enhanced the bioavailability of other compounds, making the cancer treatment more effective [2]. The results from this study support that idea.

A pattern seen across all of the data was seen in the 10 µmol piperine, 100 µmol EGCG, 10 µmol piperine, and 18.75 µmol EGCG, and 1.25 µmol Piperine groups showed an average shrinkage in the planaria length by over 10 percent on day 7 when compared to the original day one length. This shrinkage can be attributed to extreme environmental stress due to potentially high concentrations of these compounds. Similar extreme environmental stress cases involving Planaria have been found to inhibit protein expression in adult stem cells and block ion pathways [3]. This may be a problem when using the two compounds as a translational therapy for cancer cells, as environmental stress would also cause the healthy pluripotent adult cells to undergo structural changes.

Other variables that may have influenced the results of this study include varying feeding patterns, as some planaria may not have fed during their feeding times. Another variable that may have influenced the study is water levels. Despite replenishing water daily, water levels could not be maintained during the weekend. This likely could have changed the concentration levels in the solution, as less solvent would lead to a higher concentration. If the solution became too hypertonic, this could cause the planaria to shrivel.

In addition, it is also difficult to determine whether the apoptotic effect or the prevention of regeneration caused the change in the growth of planaria. However, it was theorized that EGCG and piperine would only block regeneration and not cause shrinkage. The results show otherwise. The data indicates that when the EGCG to piperine ratio was high, this led to a shrinkage in the planaria, while when the ratio was more balanced, prevention of cell proliferation was seen. This could relate to EGCG and piperine having an apoptotic effect, which is common for many cancer treatments on the market.

The data collected in this study supports the idea that EGCG and piperine both have regenerative-inhibiting properties that could be tested on cancer. A combined treatment could enhance the prevention of cell proliferation in planarian stem cells. Findings from prior studies back up the data.

In the future, the hypothesis should be tested with a cancer cell culture to identify the true effects of the two compounds on cancer cells. This change not only tests the effectiveness of piperine and EGCG, but it can also draw similarities between the behavior of planarian regeneration and cancer growth. This future study can also further examine the specific pathways piperine and EGCG affect when limiting cancer growth. This knowledge can further expand the knowledge base on how piperine and EGCG truly affect cancer cells. If this study were to be conducted with planaria, it should be repeated using more planaria to establish the conclusions further. Overall, the conclusions drawn from this study will be utilized to improve and optimize cancer treatment further and make a step forward in the affordability and efficacy of cancer treatment.